\documentclass{article}

\renewcommand{\title}[1]{
\begin{center} \Large \bf #1 \end{center}
}

\renewcommand{\author}[3]{
 \begin{center} #1 \\
  {\it #2} \\
  {\small E-mail: \texttt{#3}}
 \end{center}
\addvspace{\baselineskip}
}

\usepackage{amssymb}
\usepackage{amsmath}

\usepackage{theorem}
\theoremstyle{break}
\newtheorem{theorem}{Theorem}
\theoremstyle{break}

\theoremstyle{break}

\font\mybb=msbm10 at 12pt
\def\bb#1{\hbox{\mybb#1}}

\font\mybb=msbm10 at 12pt

\begin{document}

\begin{titlepage}

\baselineskip 5mm

\begin{flushleft}
December 2003
\end{flushleft}

\begin{flushright}
KSTS/RR-03/004\\
OCHA-PP-210 \ \ \ \ \ 
\end{flushright}

\title{Ring Structure of SUSY $*$ Product\\
and $1/2$ SUSY Wess-Zumino Model}

\author
{Akifumi Sako${}^{\dagger}$ and Toshiya Suzuki${}^{\ast} {}^{\S}$ \\ \ }
{${}^{\dagger}$ Department of Mathematics, Faculty of Science and
 Technology, Keio University\\
3-14-1 Hiyoshi, Kohoku-ku, Yokohama 223-8522, Japan\\ \ \\
${}^{\ast}$ Department of Physics, Faculty of Science, Ochanomizu University\\
${}^{\S}$ Institute of Humanities and Sciences, Ochanomizu University\\
2-1-1 Otsuka, Bunkyo-ku, Tokyo 112-8610, Japan\\ \ }
{${}^{\dagger}$ sako@math.keio.ac.jp\\
 \ \ \ \ \ \ \ \ \ \ ${}^{\ast}$ tsuzuki@phys.ocha.ac.jp}

\vspace{1cm}

\abstract{ 
There are two types of non(anti-)commutative deformation of 
D=4, N=1 supersymmetric field theories and D=2, N=2 theories. 
One is based on the non-supersymmetric star product and 
the other is based on the supersymmetric star product .
These deformations cause partial breaking of supersymmetry in general.  
In case of supersymmetric star product, the chirality is broken by the
 effect of the supersymmetric star product, 
then it is not clear that lagrangian or observables including F-terms
preserve part of supersymmetry.
In this article, we investigate the ring structure whose product is
 defined by the supersymmetric star product.
We find the ring whose elements correspond to $1/2$ SUSY F-terms. Using
 this, the $1/2$ SUSY invariance of the Wess-Zumino model is
 shown easily and directly.
}

\vspace{10mm}

\begin{flushleft}
{\bf keywords}: non(anti-)commutative deformation, supersymmetric $*$
 product, ring structure, $1/2$ SUSY
\end{flushleft}

\end{titlepage}

%

Recently, non(anti-)commutative superspaces have attracted much
interest \cite{FerraraO} - \cite{ST}.
Today's activities in this area are strongly motivated by 
several aspects of the superstring theory, or they occur in connections
between supersymmetric field theories and supermatrix models
\cite{Seiberg},\cite{CP}-\cite{ST}.
Independently of above motivations,
from the view point of both theoretical physics and mathematics
non(anti-)commutative superspaces are quite interesting subjects.
As in cases of ordinary bosonic noncommutative field theories,
non(anti-)commutative deformations are implemented by star products \cite{DN}.
Today, it is known that there are two types of deformations which lead
us to non(anti-)commutative superspace: 
(i) one is based on the non-supersymmetric star product defined by
using the supersymmetry generator $Q$;
(ii) the other is on the supersymmetric star product defined in terms of
the covariant derivative $D$ \cite{Ferrara}. 
(In this article, 
the symbol $\star$ denotes the non-SUSY star product and
the symbol $*$ denotes the SUSY star product.
Concrete definitions of them will appear below.)

For $D=4 , N=1$ superspace, as well as for $D=2 , N=2$ one, 
we introduce (anti-)chiral superfields.
In usual (anti-)commutative supersymmetric field theories, 
using (anti-)chiral superfields 
we construct a Wess-Zumino model, which is invariant under the
supersymmetry transformation, from F-terms and D-terms.
In \cite{Seiberg}, Seiberg considered the deformation of type (i), and showed
that the superalgebra is deformed and $1/2$ SUSY survive.
(Obeying Seiberg \cite{Seiberg}, we call $1/2$ SUSY as a half
 of supersymmetry generated by either $Q$ or $\bar{Q}$.)
By definition of the $\star$ product if $A$ and $B$ are (anti-)chiral
superfield $A \star B$ is trivially (anti-)chiral, 
in short the $\star$ product does not break (anti-)chirality.
As a result, the F-term is invariant under the surviving half supersymmetry. 
After all, the Wess-Zumino lagrangian is invariant under $1/2$ SUSY.
On the other hand, in \cite{Ferrara}, Ferrara {\it et al.} investigated
effects of the type (ii) deformation. 
There, the scenario of survival of supersymmetry is different from the
one in the case of type (i). 
In contrast to the case (i), the superalgebra is not deformed. 
However the $*$ product does not preserve (anti-)chirality. As a
result, the F-terms break fractions of supersymmetry.
They performed explicit calculations for Wess-Zumino models with up to
degree 5 superpotential and showed that the F-term breaks $1/2$ SUSY and
the total lagrangian is invariant under the rest of SUSY.

{}From these explicit examples of Wess-Zumino models preserving the
$1/2$ SUSY, 
it is natural to ask whether fractions of the supersymmetry survive for other superpotentials (more than degree 5) 
and what observables are invariant under the $1/2$ SUSY.
For Euclidean case we find the $1/2$ SUSY of them by using the result of
the non-SUSY $\star$ product theory.
But it is not direct way and it does not work for Minkowski case.
Therefore it is important subject to understand $1/2$ SUSY in terms of
the SUSY $*$ product.  
In the SUSY $*$ product theory, the (anti-)chirality is broken by the
effect of the SUSY $*$ product deformation.
This breaking makes it difficult to see the $1/2$ SUSY invariance.
The aim of this letter is to solve this problem and to provide a way of
understanding $1/2$ SUSY in the framework of SUSY $*$ theory.   

We constrain our analysis to a rather simple case.
We concentrate on models constructed from a single chiral scalar superfield
$\Phi$, which carries no flavor nor color. 
We find a set of superfields constructed from the chiral scalar superfield and
its covariant derivatives.
This set has two important properties:
(a) it is closed under the $*$ product, i.e. it is a ring defined
by using the $*$ product; (b) $\int d \theta^2$ of
its element is invariant under the $1/2$ SUSY.
As an example, we show that $\Phi^n_* = \overbrace{\Phi * \cdots *
\Phi}^n$ belongs to the ring. As a consequence,
F-terms constructed of $\Phi^n_*$ preserve $1/2$ SUSY.

In both (i) and (ii) cases, 
deformation parameters look like breaking explicit Lorenz invariance. 
In \cite{Seiberg}, however, it was shown that the
deformation of the Wess-Zumino lagrangian is Lorentz invariant,
that is, the deformation parameters appear in the lagrangian through
Lorentz invariant combinations. 
We also check this statement in a bit different way from one in
\cite{Seiberg}.

%
First of all, we present some formulas, which are useful for the $D=4 , N=1$
superspace calculation and for the $D=2 , N=2$ one. 
Spacetime signature is chosen as Minkowski type in the following
expression, but most parts of this article are also valid for Euclid
space. Only the [Th\ref{ftermsusy}] is restricted to the case of Euclid
space, to maintain the hermiticity of the lagrangian. 

We start with the $D=4 , N=1$ case. We use conventions of
\cite{WessBagger}.

The covariant derivatives $D_{\alpha}$ and $\bar{D}_{\dot{\alpha}}$ are
defined by
\begin{eqnarray}
D_{\alpha} &=& \ \ \partial_{\theta^{\alpha}} + i \sigma^{\mu}_{\alpha \dot{\alpha}} \bar{\theta}^{\dot{\alpha}} \partial_{\mu} \ , \nonumber \\
\bar{D}_{\dot{\alpha}} &=& - \partial_{\bar{\theta}^{\dot{\alpha}}} - i \theta^{\alpha} \sigma^{\mu}_{\alpha \dot{\alpha}} \partial_{\mu} \ .
\label{D4}
\end{eqnarray}
They satisfy the following anti-commutation relations:
\begin{equation}
\{ D_{\alpha} , \bar{D}_{\dot{\alpha}} \} = -2i \sigma^{\mu}_{\alpha \dot{\alpha}} \partial_{\mu} \ , \ others=0 \ .
\label{DCR4}
\end{equation}

The supersymmetry generators $Q_{\alpha}$ and $\bar{Q}_{\dot{\alpha}}$ are
constructed so that they anti-commute with the covariant derivatives
$D_\alpha$ and $\bar{D}_{\dot{\alpha}}$. Their explicit forms are given as
\begin{eqnarray}
Q_{\alpha} &=& \ \ \partial_{\theta^{\alpha}} - i \sigma^{\mu}_{\alpha \dot{\alpha}} \bar{\theta}^{\dot{\alpha}} \partial_{\mu} \ , \nonumber \\
\bar{Q}_{\dot{\alpha}} &=& - \partial_{\bar{\theta}^{\dot{\alpha}}} + i \theta^{\alpha} \sigma^{\mu}_{\alpha \dot{\alpha}} \partial_{\mu} \ .
\label{Q4}
\end{eqnarray}
Their anti-commutators produce bosonic translation operators $\partial_\mu$:
\begin{eqnarray}
\{ Q_{\alpha} , \bar{Q}_{\dot{\alpha}} \} = 2i \sigma^{\mu}_{\alpha \dot{\alpha}} \partial_{\mu} \ , \ others=0 \ .
\label{QCR4}
\end{eqnarray}

Using the covariant derivatives, we define (anti-)chiral
superfields.
The chiral superfields are superfields which are constrained by the following condition,
\begin{equation}
\bar{D}_{\dot{\alpha}} \Phi = 0 \ .
\label{DPhi4}
\end{equation}
In a similar way, we also define the anti-chiral superfields as
\begin{equation}
D_{\alpha} \bar{\Phi} = 0 \ . 
\label{DbPhi4}
\end{equation}
An important property of the (anti-)chiral superfield, which will be
used in the following, is that its highest component is invariant
under a half of supersymmetry:
\begin{equation}
Q D^2 \Phi |_{\bar{\theta}= 0} = 0 \ , \ \bar{Q} \bar{D}^2 \bar{\Phi} |_{\theta = 0} = 0 \ ,
\label{qd4}
\end{equation}
and becomes some total derivative under the rest:
\begin{equation}
\bar{Q} D^2 \Phi |_{\bar{\theta}= 0} = total \ derivative \ , \ Q \bar{D}^2 \bar{\Phi} |_{\theta = 0} = total \ derivative \ .
\end{equation}

Now we turn to the $D=2 , N=2$ case. The superspace coordinates are
$(z,\bar{z},\theta_\pm,\bar{\theta}_\pm)$. All of them can be considered
as independent variables.
The covariant derivatives $D_{\pm}$ and $\bar{D}_{\pm}$ are given as
\begin{eqnarray}
D_{+} = \partial_{\theta_{-}} - i \bar{\theta}_{-} \partial_{z} &,& D_{-} = \partial_{\theta_{+}} - i \bar{\theta}_{+} \partial_{\bar{z}} \ , \nonumber \\
\bar{D}_{+} = \partial_{\bar{\theta}_{-}} - i \theta_{-} \partial_{z} &,& \bar{D}_{-} = \partial_{\bar{\theta}_{+}} - i \theta_{+} \partial_{\bar{z}} \ ,
\label{D2}
\end{eqnarray}
whose anti-commutation relations are
\begin{equation}
\{ D_{+} , \bar{D}_{+} \} = -2i \partial_{z} \ , \ \{ D_{-} , \bar{D}_{-} \} = -2i \partial_{\bar{z}} \ , \ others=0 \ .
\label{DCR2}
\end{equation}

The supersymmetry generators $Q_{\pm}$ and $\bar{Q}_{\pm}$, and
their anti-commutation relations are given as
\begin{eqnarray}
Q_{+} = \partial_{\theta_{-}} + i \bar{\theta}_{-} \partial_{z} &,& Q_{-} = \partial_{\theta_{+}} + i \bar{\theta}_{+} \partial_{\bar{z}} \ , \nonumber \\
\bar{Q}_{+} = \partial_{\bar{\theta}_{-}} + i \theta_{-} \partial_{z} &,& \bar{Q}_{-} = \partial_{\bar{\theta}_{+}} + i \theta_{+} \partial_{\bar{z}} \ ,
\label{Q2}
\end{eqnarray}
and
\begin{equation}
\{ Q_{+} , \bar{Q}_{+} \} = +2i \partial_{z} \ , \ \{ Q_{-} , \bar{Q}_{-} \} = +2i \partial_{\bar{z}} \ , \ others=0 \ ,
\label{QCR2}
\end{equation}
respectively.

In a similar way of the $D=4 , N=1$ case, the (anti-)chiral superfields are defined by
\begin{equation}
\bar{D}_{\pm} \Phi = 0 \ ,
\label{DPhi2}
\end{equation}
and
\begin{equation}
D_{\pm} \bar{\Phi} = 0 \ . 
\label{DbPhi2}
\end{equation}
They satisfy
\begin{equation}
Q D^2 \Phi |_{\bar{\theta}= 0} = 0 \ , \ \bar{Q} \bar{D}^2 \bar{\Phi}|_{\theta = 0}  = 0 \ ,
\label{qd2}
\end{equation}
and
\begin{equation}
\bar{Q} D^2 \Phi |_{\bar{\theta}= 0} = total \ derivative \ , \ Q \bar{D}^2 \bar{\Phi} |_{\theta = 0} = total \ derivative \ .
\end{equation}

%

Let us introduce non(anti-)commutative deformation into superspaces.
Because our main purpose is to investigate the ring structure coming
from the $*$ product, it is appropriate to follow the procedure given in
\cite{Ferrara}.

Firstly, we introduce the notion of left/right covariant derivatives. 
The left covariant derivative is identical to
the ordinary covariant derivative
\begin{equation}
\overrightarrow{D} \Phi = D \Phi \ .
\label{lD}
\end{equation}
On the other hand, the right covariant derivative is defined through the
following relation
\begin{equation}
\Phi \overleftarrow{D} = (-1)^{p_{D} (p_{\Phi} + 1)} \overrightarrow{D} \Phi \ ,
\label{rD}
\end{equation}
where $p_{{\cal O}}$ denotes the parity of ${\cal O}$ : for odd quantity
$p_{{\cal O}_{odd}}=1$ and for even one $p_{{\cal O}_{even}}=0$. 
The Leibniz rules hold for both : 
\begin{eqnarray}
\overrightarrow{D} (\Phi \Psi) &=& \overrightarrow{D} (\Phi) \Psi + (-1)^{p_{D} p_{\Phi}} \Phi \overrightarrow{D} (\Psi) \ , \nonumber \\
(\Phi \Psi) \overleftarrow{D} &=& \Phi (\Psi) \overleftarrow{D} + (-1)^{p_{D} p_{\Psi}} (\Phi) \overleftarrow{D} \Psi \ .
\label{Lr}
\end{eqnarray}

In terms of the left and right covariant derivatives, we can define the 
supersymmetric Poisson bracket $\{ \ , \ \}_{1}$ 
\begin{equation}
\{ \Phi , \Psi \}_{1} = P^{\mu \nu} \partial_{\mu} \Phi \partial_{\nu} \Psi + P^{\alpha \beta} \Phi \overleftarrow{D}_{\alpha} \overrightarrow{D}_{\beta} \Psi + \partial_{\mu} \Phi P^{\mu \alpha} \overrightarrow{D}_{\alpha} \Psi + \Phi \overleftarrow{D}_{\alpha} P^{\alpha \mu} \partial_{\mu} \Psi \ ,
\label{PB-1}
\end{equation}
where $P^{\mu \nu}$ are anti-symmetric, $P^{\alpha \beta}$ symmetric,
and $P^{\mu \alpha} = - P^{\alpha \mu}$.
If we replace $D$ by $\bar{D}$ in (\ref{PB-1}), we obtain another
supersymmetric Poisson bracket $\{ \ , \ \}_{2}$
\begin{equation}
\{ \Phi , \Psi \}_{2} = P^{\mu \nu} \partial_{\mu} \Phi \partial_{\nu} \Psi + P^{\dot{\alpha} \dot{\beta}} \Phi \overleftarrow{\bar{D}}_{\dot{\alpha}} \overrightarrow{\bar{D}}_{\dot{\beta}} \Psi + \partial_{\mu} \Phi P^{\mu \dot{\alpha}} \overrightarrow{\bar{D}}_{\dot{\alpha}} \Psi + \Phi \overleftarrow{\bar{D}}_{\dot{\alpha}} P^{\dot{\alpha} \mu} \partial_{\mu} \Psi \ .
\label{PB2-2}
\end{equation}

Also, we can construct non-supersymmetric Poisson bracket $\{ \ , \
\}_{3}$ and $\{ \ , \ \}_{4}$ , using $Q$ instead of $D$:
\begin{eqnarray}
\{ \Phi , \Psi \}_{3} &=& P^{\mu \nu} \partial_{\mu} \Phi \partial_{\nu} \Psi + P^{\alpha \beta} \Phi \overleftarrow{Q}_{\alpha} \overrightarrow{Q}_{\beta} \Psi + \partial_{\mu} \Phi P^{\mu \alpha} \overrightarrow{Q}_{\alpha} \Psi + \Phi \overleftarrow{Q}_{\alpha} P^{\alpha \mu} \partial_{\mu} \Psi \ ,
\label{PB-3} \\
\{ \Phi , \Psi \}_{4} &=& P^{\mu \nu} \partial_{\mu} \Phi \partial_{\nu} \Psi + P^{\dot{\alpha} \dot{\beta}} \Phi \overleftarrow{\bar{Q}}_{\dot{\alpha}} \overrightarrow{\bar{Q}}_{\dot{\beta}} \Psi + \partial_{\mu} \Phi P^{\mu \dot{\alpha}} \overrightarrow{\bar{Q}}_{\dot{\alpha}} \Psi + \Phi \overleftarrow{\bar{Q}}_{\dot{\alpha}} P^{\dot{\alpha} \mu} \partial_{\mu} \Psi \ .
\label{PB-4}
\end{eqnarray}

Now, we define the supersymmetric star product, using the supersymmetric
Poisson bracket of type $\{ \ , \ \}_{1}$.
The SUSY star product ($*$) is defined by
\begin{equation}
\Phi * \Psi = e^{P}(\Psi , \Phi) = \sum_{n=0}^{\infty} \frac{1}{n !} P^{n}(\Phi , \Psi) \ ,
\label{SP}
\end{equation}
where
\begin{equation}
P^{n}(\Phi , \Psi) = \sum_{A_{1},...,A_{n};B_{1},...B_{n}}
 (-1)^{\rho_{A_{1},...,A_{n}}^{B_{1},...,B_{n}}} \Phi
 \overleftarrow{D}_{A_{1}} ... \overleftarrow{D}_{A_{n}} P^{A_{1} B_{1}}
 ... P^{A_{n} B_{n}} \overrightarrow{D}_{B_{n}}
 ... \overrightarrow{D}_{B_{1}} \Psi \ ,
\label{Pn}
\end{equation}
and
\begin{equation}
\rho_{A_{1},...,A_{n}}^{B_{1},...,B_{n}} = \sum_{i=1}^{n-1} (p_{A_{i}} + p_{B_{i}}) \sum_{j=i+1}^{n} p_{A_{j}} \ .
\label{rho}
\end{equation}
Notice that non(anti-)commutative parameters $P^{A B}$ are covariantly
constant, that is, $D_{C} P^{A B} = 0$.
One can show the associativity of the star product
\begin{equation}
\Phi * (\Psi * X) = (\Phi * \Psi) * X \ .
\label{**}
\end{equation}

Replacing $D$ by $Q$ in Eq. (\ref{Pn}) makes another star product.  
This kind of star product is called the non-SUSY star product.
We use the symbol $\star$ to denote the non-SUSY star product. It is
equivalent to the star product investigated in \cite{Seiberg}.


Let us consider the $D=4 , N=1$ case. We do not deal with it for general
non(anti-)commutative parameter $P^{AB}$. Instead, we constrain our
analysis to a special case:
\begin{equation}
P^{\alpha \beta} \neq 0 \ , \ others=0 \ .
\label{P4}
\end{equation}
With this setting, we obtain
\begin{eqnarray}
\Phi * \Psi &=& \Phi \cdot \Psi + P^{\alpha \beta} \Phi \overleftarrow{D}_{\alpha} \overrightarrow{D}_{\beta} \Psi + \frac{1}{4} det P \Phi \overleftarrow{D}^{2} \overrightarrow{D}^{2} \Psi \nonumber \\
 &=& \Phi \cdot \Psi + (-)^{(p_\Phi + 1)} P^{\alpha \beta} D_{\alpha} \Phi D_{\beta} \Psi - \frac{1}{4} det P D^{2} \Phi D^{2} \Psi \ ,
\label{SP4}
\end{eqnarray}
where $\overleftarrow{D}^{2} = \overleftarrow{D}_\alpha
\overleftarrow{D}_\beta \epsilon^{\alpha \beta}$ and
$\overrightarrow{D}^{2} = \epsilon^{\beta \alpha} \overrightarrow{D}_\alpha
\overrightarrow{D}_\beta$ .
Remark that the $P$ expansion of the $*$ product terminates at finite
order due to the fact that $D_\alpha D_\beta D_\gamma=0$.

In the $D=2 , N=2$ case,
we use the same setting as (\ref{P4}), 
so we obtain the same $P$ expansion of the $*$ product (\ref{SP4}).
(Indices $\alpha,\beta,...$ take $+$ or $-$.)

%

Now preparation is finished, so let us start to investigate the ring structure and 
$1/2$ SUSY of F-terms, and to analyze Lorentz invariance of them.
In the following of this article,
 we treat only the case where $1/2$ SUSY is generated by $Q$. 
(We can discuss the other case, that is, where $1/2$ SUSY corresponds to
$\bar{Q}$ by a similar way.)
%
We concentrate on simple models which are constructed from a single
chiral scalar superfield $\Phi$ with no flavor nor color.
So the identity $D_\alpha \Phi D_\beta \Phi D_\gamma \Phi =
0$ holds, which makes our analysis easy.
%
Since the structure of the $*$ product is the same for both the $D=4 ,
N=1$ case and the $D=2 , N=2$ case, which is given by (\ref{SP4}),  
the analysis can be performed in a uniform way.
Therefore, statements given in the following are valid for both cases.

Let us suppose following three types of set 
$X=\{D^\alpha \Phi D_\alpha \Phi\}$, 
$Y=\{\Phi\}$, $Z=\{D^2 \Phi^n | n=1,2,\cdots\}$,
and let $x$, $y$ and $z_i$ be their elements, i.e. $x \in
X$, $y \in Y$ and $z_i \in Z$, where $i=1,2,\cdots $.
We introduce following three types of polynomial ring
\begin{eqnarray}
{\bb R}[X] &=& \{ \sum_{k} a_k (D^\alpha \Phi D_\alpha \Phi)^k \}
          = \{a_1 + a_2 D^\alpha \Phi D_\alpha \Phi \; \} \ , \\
{\bb R}[Y] &=& \{ \sum_{k} a_k \Phi^k \} \ , \\
{\bb R}[Z] &=& \{ \sum_N \sum_{\{{k_1,\ldots , k_N} \}} a_{k_1,\ldots , k_N}  
\prod_i^N (z_i)^{k_i} | N \in {\bb Z}_+ \} \ ,
\end{eqnarray}
where $a_i, a_{k_1,\ldots , k_N} \in {\bb R} $. 
(One can replace real number field ${\bb R} $ by 
 arbitrary field ${\bb F}$.
This change does not affect validity of following arguments.) 
The multiplication of these polynomial rings is determined by 
ordinary multiplication.
Let $R_i(X)$,$R_i(Y)$ and
$R_i(Z)$ be polynomials belonging to the polynomial rings i.e. for an arbitrary index $i$, $R_i(X) \in {\bb R}[X]$,
$R_i(Y) \in {\bb R}[Y]$ and $R_i(Z) \in {\bb R}[Z]$.
Next step, we define some sets as follows,
\begin{eqnarray}
D_\alpha {\bb R}[X] &\equiv& \{ D_\alpha R_i(X) \; | \; \forall R_i(X) \in {\bb R}[X] \}
          = \{ a_i D_\alpha \Phi D^2 \Phi \; | \forall a_i \in {\bb R} \} \ ,
          \label{D4formura1} \\
D^2{\bb R}[X] &\equiv& 
           \{ D^2( R_i(X) ) \; | \; \forall R_i(X) \in {\bb R}[X]  \}
          = \{  a_i D^2 \Phi D^2 \Phi \; \; | \; \forall a_i \in {\bb R} \}
          \nonumber \\
          & \subset & {\bb R}[Z] \ , \\
D_\alpha {\bb R}[Y] &\equiv& \{ D_\alpha R_i(Y) \; | \; \forall R_i(Y) \in {\bb R}[Y]  \}
              =\{D_\alpha \Phi \frac{\delta R_i(Y )}{\delta \Phi}
                \; | \; \forall R_i(Y) \in {\bb R}[Y] \}
              \nonumber \\
              &\subset& \{ (D_\alpha \Phi ) R_i(Y) 
               \; | \; \forall R_i(Y) \in {\bb R}[Y] \} \ , \\
D^2{\bb R}[Y] &\equiv& 
              \{ D^2( R_i(Y) ) \; | \; \forall R_i(Y) \in {\bb R}[Y]  \}
              \nonumber \\
              &\subset& 
              \{ \sum (a_{ij}R_i(Y)R_j(Z)+ b_{ij}R_i(X)R_j(Y)) \; | 
              \; a_{ij} , b_{ij}\in {\bb R} \} \ , \\
D_\alpha {\bb R}[Z] &\equiv& \{ D_\alpha R_i(Z) \; | 
                  \; \forall R_i(Z) \in {\bb R}[Z]  \} =\{ 0 \} \ , \\
D^2 {\bb R}[Z] &\equiv& \{ D^2 R_i(Z) \; | 
                  \; \forall R_i(Z) \in {\bb R}[Z]  \}= \{ 0 \} \ . 
                  \label{D4formura6}
\end{eqnarray}

We define ${\bb R}[XYZ]$ by the set of all polynomials that are produced
by the elements of $X$, $Y$ and $Z$:
\begin{eqnarray}
{\bb R}[XYZ]&\equiv & \Big\{ \sum_{ijk} R_i(X)R_j(Y)R_k(Z)
                  \; | \; R_i(X) \in{\bb R}[X],  \;
                  R_j(Y)\in {\bb R}[Y], \; R_k(Z)\in {\bb R}[Z] \Big\}
 \nonumber \\
 &=& \Big\{ (D^\alpha \Phi D_\alpha \Phi) \sum_N \sum a_{k,k_1,\ldots ,k_N } \Phi^{k} (D^2 \Phi)^{k_1} \cdots (D^2 \Phi^N)^{k_N} \nonumber \\
& & + \sum_N \sum b_{k,k_1,\ldots ,k_N }\Phi^{k} (D^2 \Phi)^{k_1} \cdots (D^2 \Phi^N)^{k_N} \Big\} ,
\end{eqnarray}
where $a_{k,k_1,\ldots ,k_N }$ and $b_{k,k_1,\ldots ,k_N }$ are C-number coefficients.

Note that ${\bb R}[XYZ]$ is a polynomial ring produced by the 
elements of $X$,
$Y$ and $Z$ whose product is defined by ordinary multiplication.\\

Let us prove the following theorem that is a key to understand the 
$1/2$ SUSY invariance of the Wess-Zumino action.

\begin{theorem}[$*$ Ring] \label{ring}
Take ${\bb R}[XYZ]$ and $*$ product as above. 
Then ${\bb R}[XYZ]$ is a polynomial ring constructed by the elements of $X$,
$Y$ and $Z$ whose product is defined by $*$ product
i.e.  if $R_1$ and $R_2$ belong to ${\bb R}[XYZ]$ then 
$R_1 \pm R_2 \in {\bb R}[XYZ]$ and $R_1 * R_2 \in {\bb R}[XYZ]$.
\end{theorem}

{\bf Proof}\\
It is enough for the proof that we show 
${\bb R}[XYZ]$ is closed under the $*$ product.
For an arbitrary element of ${\bb R}[XYZ]$,
$R_m [XYZ] \equiv \sum_{ijk} R_i(X)R_j(Y)R_k(Z) \in {\bb R}[XYZ]$, 
$\exists R_{m'} [XYZ]  \in {\bb R}[XYZ]$ that satisfies
\begin{eqnarray}
D_\alpha R_m [XYZ]&=& \sum_{ijk} \{(D_\alpha R_i(X))R_j(Y)R_k(Z)
               + R_i(X)(D_\alpha R_j(Y))R_k(Z) \nonumber \\
            {}  && \makebox{ } \hspace{7mm} {}
                 +R_i(X)R_j(Y)(D_\alpha R_k(Z)) \} \nonumber \\
             &=&  D_\alpha \Phi R_{m'} [XYZ] \ .
\label{DRl}
\end{eqnarray}
We use (\ref{D4formura1}) $\sim$ (\ref{D4formura6}) here.
Similarly, we find that $D^2 R_m[XYZ]$ belongs to ${\bb R}[XYZ]$, i.e.
\begin{equation}
D^2 R_m [XYZ] \in {\bb R}[XYZ] .  
\end{equation}
Using these results, one can show that for  
$\forall R_m[XYZ], R_n[XYZ] \in {\bb R}[XYZ]$, \\  
$ \exists R_{m'} [XYZ], R_{n'} [XYZ], R_p[XYZ], R_q[XYZ] \in {\bb R}[XYZ] $ that satisfy
\begin{eqnarray}
R_m [XYZ] * R_n [XYZ] &=& R_m [XYZ] R_n [XYZ]
\nonumber \\
&&
+ P^{\alpha \beta} R_m [XYZ] \overleftarrow{D}_\alpha \overrightarrow{D}_\beta R_n [XYZ]
\nonumber \\
&&+ \frac{1}{4} det P R_m [XYZ] \overleftarrow{D}^2 \overrightarrow{D}^2 R_n [XYZ] \nonumber \\
 &=& R_m [XYZ] R_n [XYZ] \nonumber \\
 && + P^{\alpha \beta} D_\alpha \Phi D_\beta \Phi R_{m'} [XYZ] R_{n'} [XYZ] \nonumber \\
 &&+ R_p [XYZ] R_q [XYZ] \ .
\end{eqnarray}
Because of symmetric property of $P^{\alpha \beta}$, the second term
 vanishes, i.e.
\begin{equation}
P^{\alpha \beta} D_\alpha \Phi D_\beta \Phi R_{m'} [XYZ] R_{n'} [XYZ] = 0 ,
\label{PRR}
\end{equation}
then we can conclude
$R_m [XYZ] * R_n [XYZ] \in {\bb R}[XYZ]$.  
\begin{flushright}
$\blacksquare$
\end{flushright}

Let us see the relation between ${\bb R}[XYZ]$ and 
$1/2$ SUSY, here.

\begin{theorem}[Ring and $1/2$ SUSY] \label{ringsusy}
Take ${\bb R}[XYZ]$ as above.
If $R_i [XYZ] \in {\bb R}[XYZ] $, then 
$\int d^2 \theta R_i [XYZ] $ is invariant under $1/2$
SUSY transformation. 
\end{theorem}

{\bf Proof}\\
Arbitrary $R_i [XYZ] \in {\bb R}[XYZ]$
is expressed as 
\begin{eqnarray}
R_i [XYZ]&=& \sum a_{nk} (D^\alpha \Phi D_\alpha \Phi)(\Phi)^n R_k[Z]
\nonumber \\
&&+ \sum b_{nl} (\Phi)^n R_l[Z] ,
\label{DD-1}
\end{eqnarray}
where $R_k(Z)$ and $R_l(Z)$ are elements of ${\bb R}[Z]$. 
When $D^2$ operates on the first term of Eq. (\ref{DD-1}),
\begin{eqnarray}
D^2 && \!\!\!\!\!\!\!\sum a_{nk} (D^\alpha \Phi D_\alpha \Phi)(\Phi)^n R_k[Z]
\nonumber \\
&=&-\sum a_{nk}D^2 \Phi D^2 \Phi (\Phi)^n R_k[Z]
-\sum a_{nk} n (D^2 \Phi) (D^\alpha \Phi D_\alpha \Phi) \Phi^{n-1} R_k[Z]
\nonumber \\
&=& -\sum a_{nk}\frac{1}{n+1}(D^2 \Phi) (D^2 (\Phi)^{n+1}) R_k[Z] .
\label{DD-2}
\end{eqnarray}
When $D^2$ operates on the second term of Eq. (\ref{DD-1}), 
\begin{eqnarray}
D^2 \sum b_{nl} (\Phi)^n R_l[Z] = 
\sum b_{nl} (D^2(\Phi^n)) R_l[Z] .
\label{DD-3}
\end{eqnarray}
Eqs.(\ref{DD-2}) and (\ref{DD-3}) show that
$D^2 R_i [XYZ] \in {\bb R}[Z]$.
Then some $R_j[Z](\in {\bb R}[Z])$ exists that satisfies
\begin{eqnarray}
\int d^2 \theta R_i [XYZ] = D^2 R_i [XYZ] |_{\bar{\theta}=0}
=R_j[Z]|_{\bar{\theta}=0} .
\end{eqnarray}
Recall that 
$R_j[Z]$ is some polynomial of $D^2$ exact terms and 
$D^2$ exact terms are invariant under $Q$ (see Eq.(\ref{qd4}) or Eq.(\ref{qd2})).
Therefore, it is proved that
$\int d^2 \theta R_i [XYZ] $ is invariant under $1/2$ SUSY transformation.
\begin{flushright}
$\blacksquare$
\end{flushright}

%

Using above theorems, we can easily show that $1/2$ SUSY
invariance of F-terms.

\begin{theorem}[$1/2$ SUSY invariance of F-terms] \label{ftermsusy}
Take $*$ product as above.
Let $\Phi$ be a chiral superfield.
Then, for arbitrary $n \in {\bb N}$,
$\int d^2 \theta (\Phi)_*^n =\int d^2 \theta \Phi* \cdots *\Phi $
is invariant under $1/2$ SUSY transformation. \\
\end{theorem}

{\bf Proof}\\
Because of the theorem of Ring structure [Th\ref{ring}], 
$(\Phi)_*^{n} \in {\bb R}[XYZ]$.
{}From [Th\ref{ringsusy}], it is proved that 
$\int d^2 \theta (\Phi)_*^n $ is $1/2$SUSY invariant. 
\begin{flushright}
$\blacksquare$
\end{flushright}

The F-term $1/2$ SUSY of the $*$ product is
shown by the F-term $1/2$ SUSY of the $\star$ product  
without the ring ${\bb R[XYZ]}$, as follows.
When we consider the Euclidean space, we can 
take $\bar{\theta}=0$ before $\int d\theta^2$.
Therefore replacing the $D$ operators with the $Q$ operators 
makes no difference in the computation of the F-terms.
In short, we can exchange $*$ by $\star$ in F-terms and
the F-terms are identical in both cases.
Since the $\star$ product does not break chirality,
this fact implies the $1/2$ SUSY of the $\int d^2 \theta (\Phi)^n_*$.
\footnote{This argument was taught us by the referee of Phys. Lett. B.}

In order to see other usefulness of the ring ${\bb R}[XYZ]$, let us
construct {\it non-trivial} observables which can have non-zero
v.e.v. in $1/2$ SUSY invariant phases.
Consider observables 
\begin{equation}
\int d\theta^2 D^{\alpha} * \Phi * D_{\alpha} * \Phi * (\Phi)^n_* = 
\int d\theta^2 ((D^{\alpha} * \Phi) * (D_{\alpha} * \Phi)) * (\Phi)^n_* .
\label{obs}
\end{equation}
Since $D_\alpha * \Phi = D_\alpha \Phi$ and $D^{\alpha} \Phi *
D_{\alpha}\Phi = D^{\alpha} \Phi D_{\alpha} \Phi$, the integrand of
(\ref{obs}) is equal to $(D^{\alpha} \Phi D_{\alpha}\Phi ) * (\Phi)^n_*$.
Because $(D^{\alpha} \Phi  D_{\alpha}\Phi )$ and $(\Phi)^n_* $ belong to
${\bb R}[XYZ]$, we conclude 
$\int d\theta^2 D^{\alpha} * \Phi * D_{\alpha} * \Phi * (\Phi)^n_*$ is
$1/2$ SUSY invariant by [Th\ref{ring}] and [Th\ref{ringsusy}].

In the $\star$ theory, the $1/2$ SUSY invariance of
$\int d\theta^2 D^{\alpha} \star \Phi \star D_{\alpha} \star \Phi \star
(\Phi)^n_{\star}$, the counterpart of (\ref{obs}), is not manifest,  
as they include non-chiral objects.
But, by making use of $1/2$ SUSY of (\ref{obs}),
we conclude that they are still $1/2$ SUSY invariant.
\footnote{As noted in \cite{TY}, $Q_\alpha \star \Phi$ is a chiral
superfield in the $\star$ theory. Then by using the same argument as
above $1/2$ SUSY of $\int d\theta^2 D^{\alpha} \star \Phi \star
D_{\alpha} \star \Phi \star (\Phi)^n_{\star}$ is shown. We thank the
referee of Phys. Lett. B for pointing out this.}
Thus the ring ${\bb R}[XYZ]$ provides a new method to construct $1/2$
SUSY invariant observables which explicitly break (anti-)chirality. 

We have studied the relation between the $1/2$ SUSY and the 
$*$ product above. Not only the SUSY but also the Lorentz invariance
of the theories becomes nontrivial under the non(anti-)commutative
deformation. 
The following theorem solves this problem.

\begin{theorem}[Lorentz invariance] \label{lorentz}
Let $f$ be some Lorentz invariant superfield.
We denote $f^{n}_{*}$ as $\overbrace{f*\cdots *f}^n$. 
Then some Lorentz invariant functional $g(f, (Df)^2 , D^2 f ; \det P)$
exist
, where $(Df)^2 = D^{\alpha}f D_{\alpha}f$, 
and it satisfies that
\begin{eqnarray}
f^{n}_{*}= f^n + g .
\end{eqnarray}
Here the noncommutative parameters $P^{\alpha \beta}$
dependence only appear as the $\det P$ dependence.
This fact shows that 
$f^{n}_{*}$ is Lorentz invariant.
\end{theorem}

{\bf Proof}\\
We prove this theorem by using mathematical induction as follows.\\
{\rm (i)} $n=2$; 
\begin{eqnarray}
f*f=f^2 - \frac{1}{4} (\det P) (D^2 f)^2  
\end{eqnarray}
{\rm (ii)} Suppose
\begin{eqnarray}
f^{n}_{*}= f^n + g \ .
\end{eqnarray}
Using this,
\begin{eqnarray}
f^{n+1}_{*}= f*f^n +  f*g \ .
\end{eqnarray}
The first term is rewritten as 
\begin{eqnarray}
f*f^n= f^{n+1} - \frac{1}{4} (\det P) (D^2 f) (D^2 f^n)  .
\end{eqnarray}
The second term is given as 
\begin{eqnarray}
f*g = fg + 
(-)^{p_f+1} P^{\alpha \beta} D_{\alpha}f D_{\beta} g
-\frac{1}{4} (\det P) (D^2 f) (D^2 g) .
\label{fg}
\end{eqnarray}
We can show that the second term of Eq.(\ref{fg}) vanishes, as follows.
\begin{eqnarray}
D_{\alpha} g(f, (Df)^2 , D^2 f)&=&
(D_{\alpha}f) \frac{\partial g}{\partial f}+
(D_{\alpha}(Df)^2 ) \frac{\partial g}{\partial ((Df)^2)}+
(D_{\alpha} (D^2 f)) \frac{\partial g}{\partial (D^2 f)}
\nonumber \\
&=& (D_{\alpha}f) \frac{\partial g}{\partial f} 
-\sum_{\beta} \epsilon^{\alpha \beta} (D_{\alpha} D_{\beta} f) 
D_{\alpha} f \frac{\partial g}{\partial ((Df)^2)}.
\label{Dg}
\end{eqnarray}
Here we do not sum over the index $\alpha$.
Using Eq.(\ref{Dg}) and symmetric nature of $P^{\alpha \beta}$, 
the second term of Eq.(\ref{fg}) vanishes ;
\begin{eqnarray}
\sum_{\alpha \beta}&&\!\!\!\!\!\!\!  P^{\alpha \beta} D_{\alpha}f D_{\beta} g
\nonumber \\
 &=&\sum_{\alpha \beta \gamma} \Big\{ \frac{1}{2}P^{\alpha \beta}D_{\alpha} f D_{\beta} f 
 \frac{\partial g}{\partial f} -
 \frac{1}{2}P^{\alpha \beta} (D_{\alpha}f)
 \epsilon^{\beta \gamma} (D_{\beta} D_{\gamma} f) D_{\beta} f
 \frac{\partial g}{\partial ((Df)^2)} \Big\}=0 .
\end{eqnarray}
Therefore, 
\begin{eqnarray}
f^{n+1}_*=f^{n+1} - \frac{1}{4} (\det P) (D^2 f) (D^2 f^n) 
+ fg 
-\frac{1}{4} (\det P) (D^2 f) (D^2 g)= f^{n+1}+ g',
\end{eqnarray}
which is what we want.
Here we denote $g'$ as $-\frac{1}{4} (\det P) (D^2 f) (D^2 f^n) 
+ fg 
-\frac{1}{4} (\det P) (D^2 f) (D^2 g)$.
By the principle of mathematical induction, for all $n \ge 2$ 
(and $n=1$ that is trivial case),
$f^{n}_*$ are Lorentz invariant.
\begin{flushright}
$\blacksquare$
\end{flushright}

Note that only the symmetric property of the noncommutative parameter
$P^{\alpha \beta}$ is used in this proof.
So, we can show the Lorentz invariance for other 
star products that is defined by other Poisson brackets
like the non-SUSY Poisson bracket \cite{Seiberg} . \\

It is worth while to comment here about the ring structure of Lorentz
invariant functionals.
Let ${\bb R}_L [f, (D f)^2, D^2 f; det P]$ be a set of all polynomials
of $f$, $(D f)^2$ and $D^2 f$, with additional dependence of $det P$.
Since non(anti-)commutative parameter $P^{\alpha \beta}$ appears in
elements of ${\bb R}_L [f, (D f)^2, D^2 f; det P]$ only through $det P$,
 the elements are Lorentz invariant. We can show that ${\bb R}_L
 [f, (D f)^2, D^2 f; det P]$ is a ring whose product is defined by the
 $*$ product, by a similar way to the proof of [Th \ref{ring}]. \\

Wess-Zumino models contain not only F-terms but also D-terms.
In general, D-terms are deformed by the $*$ product, too.
However, we can show the Lorentz invariance and the SUSY invariance of
the D-terms as follows.
Let K\"ahler potential $K(\Phi , \bar{\Phi})_*$ be defined as formal
power series of $\Phi$ and $\bar{\Phi}$, where the multiplication is
defined by the
SUSY $*$ product. Using the fact that $\Phi * \bar{\Phi} = \Phi \cdot
\bar{\Phi}$ and [Th\ref{lorentz}], the K\"ahler potential is rewritten as 
\begin{eqnarray}
K(\Phi , \bar{\Phi})_* &=& \sum_{i j} c_{i j} \ \Phi_*^{i} \cdot \bar{\Phi}^j \nonumber \\
 &=& \sum_{i j} c_{i j} \ \{ \Phi^i + g_i(\Phi, (D \Phi)^2, D^2 \Phi^m; det P) \} \cdot \bar{\Phi}^j \ ,
\label{Dterm}
\end{eqnarray}
where $g_i(\Phi, (D \Phi)^2, D^2 \Phi^m; det P)$ are Lorentz invariant
functionals. Eq. (\ref{Dterm}) shows the Lorentz invariance of the
D-term.
In addition, $\int d^4 \theta K(\Phi,\bar{\Phi})_*$
is always invariant under the SUSY, then we conclude
that the total Wess-Zumino lagrangian is invariant under the $1/2$ SUSY
transformation and the Lorentz one. \\

%

Finally, we summarize main conclusions.
We proved some theorems about $1/2$ SUSY for N=1 D=4 and N=2 D=2 cases.
We discovered the ring ${\bb R}[XYZ]$ whose product is defined by 
the SUSY $*$ product.
In other words, ${\bb R}[XYZ]$ is closed under the SUSY $*$ product.
The SUSY $*$ product does not preserve (anti-)chirality,
nevertheless we proved that $\int d^2 \theta$ of ${\bb R}[XYZ]$ elements 
are $1/2$ SUSY invariant.
{} From these facts, we easily 
saw that usual F-terms, which take forms of $\int d^2 \theta \Phi^n_* $,
is invariant under $1/2$ SUSY transformation in the framework of the
SUSY $*$ deformed theory.
Furthermore, ${\bb R}[XYZ]$ made the new way to construct
$1/2$ SUSY invariant observables for both SUSY $*$ formulation and non-SUSY
$\star$ formulation.
Using this method we can construct explicitly chirality broken observables as $1/2$ SUSY invariant observables, for example 
$\int d\theta^2 D^{\alpha} * \phi * D_{\alpha} * \phi * (\phi)^n_*$.
Such observables are still $1/2$ SUSY observables after replacing $*$ by 
non-SUSY $\star$ product.
In addition, we proved the Lorentz invariance of F-terms and D-terms by explicit 
calculations.  It is possible to show the Lorentz invariance by using ring structure
as similar to the proof of $1/2$ SUSY.
\\

\noindent
{\bf Acknowledgments}

\noindent
We thank the referee of Phys. Lett. B for many useful suggestions and pointing out typos in earlier version of this article.
A.Sako is supported by 21st Century COE Program at Keio University
(Integrative Mathematical Sciences: Progress in Mathematics Motivated by Natural and Social Phenomena ).

%

\end{document}